\documentclass[onecolumn,showpacs,11pt]{revtex4}
\usepackage{graphicx}
\usepackage{dcolumn}
\usepackage{bm}
\usepackage{axodraw}
\begin{document}
\newcommand{\hs}{\hspace*{0.5cm}}
\newcommand{\vs}{\vspace*{0.5cm}}
\newcommand{\be}{\begin{equation}}
\newcommand{\ee}{\end{equation}}
\newcommand{\bea}{\begin{eqnarray}}
\newcommand{\eea}{\end{eqnarray}}
\newcommand{\ben}{\begin{enumerate}}
\newcommand{\een}{\end{enumerate}}
\newcommand{\bde}{\begin{widetext}}
\newcommand{\ede}{\end{widetext}}
\newcommand{\nn}{\nonumber}
\newcommand{\crn}{\nonumber \\}
\newcommand{\Tr}{\mathrm{Tr}}
\newcommand{\non}{\nonumber}
\newcommand{\noi}{\noindent}
\newcommand{\al}{\alpha}
\newcommand{\la}{\lambda}
\newcommand{\bet}{\beta}
\newcommand{\ga}{\gamma}
\newcommand{\va}{\varphi}
\newcommand{\om}{\omega}
\newcommand{\pa}{\partial}
\newcommand{\+}{\dagger}
\newcommand{\fr}{\frac}
\newcommand{\bc}{\begin{center}}
\newcommand{\ec}{\end{center}}
\newcommand{\Ga}{\Gamma}
\newcommand{\de}{\delta}
\newcommand{\De}{\Delta}
\newcommand{\ep}{\epsilon}
\newcommand{\varep}{\varepsilon}
\newcommand{\ka}{\kappa}
\newcommand{\La}{\Lambda}
\newcommand{\si}{\sigma}
\newcommand{\Si}{\Sigma}
\newcommand{\ta}{\tau}
\newcommand{\up}{\upsilon}
\newcommand{\Up}{\Upsilon}
\newcommand{\ze}{\zeta}
\newcommand{\ps}{\psi}
\newcommand{\Ps}{\Psi}
\newcommand{\ph}{\phi}
\newcommand{\vph}{\varphi}
\newcommand{\Ph}{\Phi}
\newcommand{\Om}{\Omega}

\title{A simple model of gauged lepton and baryon charges}

\author{P. V. Dong}
\email{pvdong@iop.vast.ac.vn} \affiliation{Institute of Physics,
VAST, P. O. Box 429, Bo Ho, Hanoi 10000, Vietnam}
\author{H. N. Long}
\email{hnlong@iop.vast.ac.vn} \affiliation{Institute of Physics,
VAST, P. O. Box 429, Bo Ho, Hanoi 10000, Vietnam}

\date{\today}

\begin{abstract}

We argue that simpler fermionic contents, responsible for the
extension of the standard model with gauged lepton and baryon
charges, can be constructed by assuming existence of so-called
leptoquarks $(j,k)$ with exotic electric charges $q_j=1/2,\
q_k=-1/2$. Some new features in our model are that (i) as the
natural consequences of anomaly cancelation the right-handed
neutrinos exist, and the number of the observed fermion families
is equal to the number of the fundamental colors; (ii) although
the lepton and baryon charges are conserved, the neutrinos can
obtain small masses through the type I seesaw mechanism in
similarity to the standard context, and the baryogenesis can be
generated in several cases. They all are natural results due to
the spontaneous breaking of these charges. Some constraints on the
new physics via flavor changing and related phenomenologies such
as the stable scalar with anomalous electric charge and interested
processes at colliders are also discussed.

\end{abstract}


\maketitle

\section{\label{intro}Introduction}

It has been evidenced that the lepton charge ($L$) and the baryon
charge ($B$) belong to exact symmetries up to very high scales
much beyond TeV scale \cite{pdg}. They may be only violated at the
scales as of the grand unification theories such as that of the
type I seesaw mechanism to explain the smallness of neutrino
masses and that to describe the stability of proton in the proton
decay question. It is therefore believed that these charges are of
exact symmetries but spontaneously broken as the gauge symmetries.
The possible phenomenologies of such theories at the TeV scale
have been recently called for much attention
\cite{foot,carone,pw,pw1,chao,py}.

If the lepton and baryon charges respectively behave as their own
gauge symmetries, it is natural to extend the standard model (SM)
gauge symmetry into a larger group $G=\mathrm{SU}(3) \otimes
\mathrm{SU}(2) \otimes \mathrm{U}(1)_Y \otimes \mathrm{U}(1)_L
\otimes \mathrm{U}(1)_B$, where the last two factors are the gauge
groups of the lepton and baryon charges, respectively. The
consistent condition of the model requires that all the anomalies
associated with $G$ must be canceled. However, in the literature
the models of lepton and baryon charges gauged are actually
complicated in the fermionic contents because they include many
exotic particles as a result to cancel the anomalies. Also, the
seesaw mechanism responsible for generating the neutrino small
masses is quite complex due to the contributions of a large
spectrum of neutral leptons \cite{foot,carone,pw,pw1,chao,py}.
Taking an example, a fourth family of leptons and a fourth family
of quarks which are replications of the ordinary fermion families
including the right-handed neutrinos were introduced into the SM,
in which the fourth family exotic leptons have $L=(+/-)3$ and the
exotic quarks having $B=(+/-)1$. The sign either plus or minus
will appropriately take place depending on which chirality left or
right assigned on the fermion multiplets \cite{pw,pw1}. Let us
remind the reader that in these models the right-handed neutrinos
of all the lepton families are in general not required on the
background of the anomaly cancelation.

In the following we propose simpler fermionic contents which are
free from all the anomalies. The leptonic and baryonic anomalies
can be removed by introducing a single family of only a fermion
kind of leptoquarks with appropriate quantum numbers. The leptonic
content in our model is thus minimal since it only includes the
ordinary right-handed neutrinos, as played in the usual type I
seesaw mechanism, emerging as a natural requirement of the
gravity-anomaly cancelation. From vanishing of another anomaly,
the number of experimentally observed fermion families is found to
be related to the number of the fundamental colors, which is equal
to three. Advantages of the model such as simplicity in the seesaw
mechanism responsible for the neutrino masses, the generation of
baryon number asymmetry, phenomenologies associated with the
leptoquarks, new gauge bosons, stable anomalously-charged scalar
as well as some constraints on the model via flavor changing are
also shown.

The rest of this article is organized as follows. In the next
section, Sec. \ref{model}, we construct the model by stressing on
new fermionic contents. Section \ref{yk} is devoted to scalar
particles, Yukawa interactions, fermion masses and some remarkable
phenomenologies. In Sec. \ref{newp}, we present constraints on the
new physics such as flavor changing neutral currents (FCNCs) in
the quark and lepton sectors, stable scalar, baryogenesis, and
effects of the new particles as well as their possible processes
and productions at the existing and future colliders. We summarize
our results and make conclusions in the last section---Sec.
\ref{conclus}.

\section{\label{model}The model}

To cancel the anomalies associated with the gauge symmetry
$G=\mathrm{SU}(3) \otimes \mathrm{SU}(2) \otimes \mathrm{U}(1)_Y
\otimes \mathrm{U}(1)_L \otimes \mathrm{U}(1)_B$, we introduce
into the SM particle content a single kind of only
colored-fermions $j$ and $k$ with $L=-1$, $B=-1$ and electric
charges $q_j=1/2$, $q_k=-1/2$, called leptoquarks. In addition,
the right-handed neutrinos corresponding to the SM lepton families
will be included in order to cancel the gravity anomaly. The
particle content in our model under $G$ transform as \bc
\begin{tabular}{|l|l|}
\hline

SM leptons & $\psi_{aL}\equiv(\nu_{aL}\ e_{aL})\sim (1,2,-1/2,1,0)$ \\

& $e_{aR}\sim (1,1,-1,1,0)$ \\ \hline

 & $Q_{aL}\equiv(u_{aL}\ d_{aL})\sim (3,2,1/6,0,1/3)$ \\

SM quarks & $u_{aR}\sim (3,1,2/3,0,1/3)$ \\

& $d_{aR}\sim (3,1,-1/3,0,1/3)$
\\ \hline

Right-handed neutrinos & $\nu_{aR}\sim (1,1,0,1,0)$\\ \hline

 & $F_L\equiv(j_L\ k_L)\sim (3,2,0,-1,-1)$\\

Leptoquarks & $j_R\sim (3,1,1/2,-1,-1)$\\

&  $k_R\sim (3,1,-1/2,-1,-1)$
\\ \hline
\end{tabular}\ec
where $a=1,2,3$ is family index, and the values in the parentheses
denote quantum numbers based on the $(\mathrm{SU}(3),
\mathrm{SU}(2),\mathrm{U}(1)_Y, \mathrm{U}(1)_L, \mathrm{U}(1)_B)$
symmetries, respectively.

One can check that all the anomalies are canceled, for examples:
\bea && [\mathrm{SU}(3)]^2 \mathrm{U}(1)_Y =0,\hs
[\mathrm{SU}(2)]^2 \mathrm{U}(1)_Y =0, \hs [\mathrm{Gravity}]^2
\mathrm{U}(1)_Y =0,\\ && [\mathrm{U}(1)_Y]^3 =0,\hs
[\mathrm{SU}(3)]^2 \mathrm{U}(1)_L =0,\hs [\mathrm{SU}(3)]^2
\mathrm{U}(1)_B =0,\\ && [\mathrm{SU}(2)]^2 \mathrm{U}(1)_L =
(3\sim \mathrm{number\ of\ SM\ lepton\ families})\times(1)\crn
&&\hspace{3cm}+(3\sim \mathrm{number\ of\ fundamental\
colors})\times(-1)=0,\\ && [\mathrm{SU}(2)]^2 \mathrm{U}(1)_B =
(3\sim \mathrm{number\ of\ SM\ quark\ families})\crn
&&\hspace{3cm}\times(3\sim \mathrm{number\ of\ fundamental\
colors})\times(1/3)\crn &&\hspace{3cm}+(3\sim \mathrm{number\ of\
fundamental\ colors})\times(-1)=0,\\ && [\mathrm{U}(1)_Y]^2
\mathrm{U}(1)_L
=3\times(-1/2)^2\times(1+1)-3\times(-1)^2\times1+3\times(0)^2\times(-1-1)\crn
&&\hspace{3cm}-3\times(1/2)^2\times(-1)-3\times(-1/2)^2\times(-1)=0,\\
&& [\mathrm{U}(1)_Y]^2 \mathrm{U}(1)_B
=3\times3\times(1/6)^2\times(1/3+1/3)-3\times3\times(2/3)^2\times(1/3)\crn
&&\hspace{3cm}-3\times3\times(-1/3)^2\times(1/3)+3\times(0)^2\times(-1-1)\crn
&&\hspace{3cm}-3\times(1/2)^2\times(-1)-3\times(-1/2)^2\times(-1)=0,\\
&&[\mathrm{Gravity}]^2\mathrm{U}(1)_L=3\times(1\sim\mbox{lepton-charge\
of\ left-handed\ neutrino})\crn
&&\hspace{3cm}-3\times(1\sim\mbox{lepton-charge\ of\ right-handed\
neutrino})=0.\eea The last equation shows that the right-handed
neutrinos are required. If one supposes that the leptoquarks have
no exotic lepton charges as we took $L=-1$, then the number of
ordinary lepton families is equal to the number of the fundamental
color (as we can see above from the $[\mathrm{SU}(2)]^2
\mathrm{U}(1)_L$ anomaly cancelation). Otherwise, the family
number will be a multiple of the color number. The similar one
with the baryonic anomaly implies that the number of the observed
quark families is also related to the color number. The
leptoquarks possessing $B=-1$ as we put are in order to cancel the
baryonic anomalies. It is noteworthy that the $[\mathrm{SU}(2)]^2
\mathrm{U}(1)_Y$ anomaly was removed because of $Y(F_L)=0$, and
this is why we assumed $q_j=-q_k=1/2$. In summary, the baryonic
and leptonic anomalies as stored in the SM particle content are
all canceled due to the presence of just leptoquarks $(j,k)$.

It is also noted that we can have another fermion content if one
reverses the chirality of leptoquarks, simultaneously changes
their sign of baryon and lepton charges ($L=+1$, $B=+1$):
$F_R=(j_R\ k_R)\sim (3,2,0,1,1)$, $j_L\sim (3,1,1/2,1,1)$, and
$k_L\sim (3,1,-1/2,1,1)$. In the following, we consider the case
with the first fermionic content as given in the table above. The
scalar sector will be introduced as usual to generate the mass for
the particles and to make the model viable.

\section{\label{yk}Scalars, Yukawa interactions and related phenomena}

The Higgs doublet of the SM transforms as follows $H=(H^+\
H^0)\sim (1,2,1/2,0,0)$. We see that $H$ does not couple the
leptoquarks and the ordinary quarks together because of the
$\mathrm{U}(1)_{Y,L,B}$ gauge invariance. The Yukawa interactions
are \bea -\mathcal{L}_{H}&=&h^e_{ab}\bar{\psi}_{aL} H
e_{bR}+h^\nu_{ab}\bar{\psi}_{aL} i\sigma_2H^* \nu_{bR} \crn &&
+h^d_{ab} \bar{Q}_{aL} H d_{bR} +h^u_{ab} \bar{Q}_{aL} i\sigma_2
H^* u_{bR}\crn && + h^k \bar{F}_L H k_R +h^j \bar{F}_L i\sigma_2
H^* j_R\crn && + h.c. \eea One can check that all the ordinary
fermions, charged leptons and quarks, gain masses in the same with
the SM. The neutrinos $\nu_L$ and $\nu_R$ at this step obtain
Dirac mass as follows\be M_D=h^\nu \langle H^0 \rangle, \ee where
$\langle H^0 \rangle=174\ \mathrm{GeV}$ is the vacuum expectation
value (VEV) of $H^0$, and $M_D$ is a $3\times 3$ matrix in $a$ and
$b$ family indices. It is easily checked that the leptoquarks do
not mix with the ordinary quarks, and have masses proportional to
$\langle H^0 \rangle$. They should be uncharacteristically heavy
like top quark, but different from the top quark because these
particles possess the unusual charges such as lepton, baryon and
electricity.

Next, let us introduce a scalar singlet $S_L\sim (1,1,0,-2,0)$
which couples to the right-handed neutrinos $\nu_R$. The Yukawa
interaction is then \be -\mathcal{L}_{S_L} = \fr 1 2 \la^\nu_{ab}
\bar{\nu}^c_{aR}\nu_{bR} S_L+h.c. \ee When $S_L$ develops a VEV,
it provides not only Majorana mass for $\nu_R$:\be
M_R=\la^\nu\langle S_L \rangle,\ee which is also a $3\times 3$
matrix in family indices, but also a necessary mass for the
$\mathrm{U}(1)_L$ gauge boson $Z'_L$: $M_{Z'_L}=g_L\langle S_L
\rangle$, where $g_L$ is the $\mathrm{U}(1)_L$ gauge coupling
constant. Let us note that the lepton charge is spontaneously
broken by the $S_L$ scalar. To be consistent with the effective
theory, we should impose $\langle S_L \rangle\gg \langle H_0
\rangle$.

An interesting result from our proposal is that three active
neutrinos $(\sim \nu_L)$ gain masses via a type I seesaw mechanism
similar to the simplest seesaw extension of the SM: \be
M^{\mathrm{eff}}_\nu=-M_DM^{-1}_RM^T_D,\ee which is quite
different from the previous proposals \cite{pw,chao}. This is also
a simple one like the fermionic content in our model.

To avoid having stable colored particles we will introduce the
following scalar doublets charged under $\mathrm{U}(1)_{L,B}$ that
couples the leptoquarks to the ordinary quarks: \be
\phi=(\phi^{+7/6}\ \phi^{+1/6})\sim (1,2,2/3,1,4/3),\hs
\chi=(\chi^{+1/6}\ \chi^{-5/6})\sim (1,2,-1/3,1,4/3).\ee The
Yukawa interactions are then \bea -\mathcal{L}_{\phi\chi}&=&h'^k_a
\bar{Q}_{aL} \phi k_R +h'^u \bar{F}_L i \sigma_2\phi^*
u_{aR}+h'^j_a \bar{Q}_{aL} \chi j_R + h'^d_a \bar{F}_L i
\sigma_2\chi^* d_{aR} + h.c.\eea The leptoquarks can now decay
into a scalar and SM ordinary quark. It is noteworthy that in our
model the $\phi$ and $\chi$ scalars could not develop a VEV
because of electric charge conservation, which is also a new
feature and in contradiction to \cite{pw}. Consequently, the
leptoquarks do not mix in mass with the ordinary quarks (notice
also that they are different in electric charges), and the FCNCs
at the tree level never appear.

Let us note that the scalars $\chi$ and $\phi$ have the anomalous
fractional electric charges, unlike the ordinary quarks. With the
electric charge conservation, this implies that at least one of
the scalars is absolutely stable. Unfortunately, it maynot be a
dark matter since it has a electric charge (see also \cite{cdm,cdm1} for similar matters).

Finally, the baryon charge should be spontaneously broken too.
This may be achieved by another scalar singlet $S_B$ charged under
$\mathrm{U}(1)_B$. Then, the $\mathrm{U}(1)_B$ gauge boson $Z'_B$
will gain a mass proportional to $\langle S_B \rangle$. The
consistent condition with the effective theory also implies
$\langle S_B \rangle\gg \langle H^0 \rangle$.

The scalar potential that consists of doublets $H$, $\phi$, $\chi$
and singlets $S_L$, $S_B$ can be easily written. Here we notice
that all the charged scalars as contained in the doublets do not
mix among them and with the others, and by themselves become mass
eigenstates.

\section{\label{newp}Constraints on the new physics}

\subsection{Flavor changing}

First we consider the flavor violation in the quark sector. As
mentioned above, there is no FCNC at the tree level because of the
gauge symmetry. We also know that the ordinary quarks have
couplings to the leptoquarks. This will lead to the FCNC processes
in the ordinary quark sector at one loop level via exchange of the
leptoquarks. In this work, we consider the decay $b\rightarrow
s\gamma$ and processes associated with $K^0-\bar{K}^0$ mixing (the
mixings $D^0-\bar{D}^0$, $B^0-\bar{B}^0$ and decays $s\rightarrow
d \gamma$, $b\rightarrow d\gamma$, $t\rightarrow c \gamma$ and so
on can be similarly calculated).

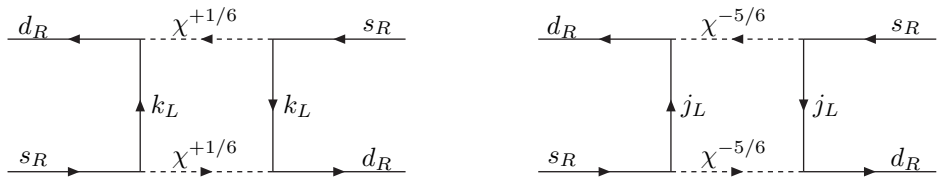
\begin{figure}[h] \bc
\begin{picture}(400,100)(0,0)
\ArrowLine(50,25)(100,25)\ArrowLine(100,25)(100,75)\ArrowLine(100,75)(50,75)
\DashArrowLine(100,25)(150,25){2}\DashArrowLine(150,75)(100,75){2}
\ArrowLine(150,25)(200,25)\ArrowLine(150,75)(150,25)\ArrowLine(200,75)(150,75)
\Text(110,50)[]{$k_L$}\Text(160,50)[]{$k_L$} \Text(60,30)[]{$s_R$}
\Text(60,80)[]{$d_R$}\Text(190,80)[]{$s_R$} \Text(190,30)[]{$d_R$}
\Text(125,82)[]{$\chi^{+1/6}$} \Text(125,32)[]{$\chi^{+1/6}$}


\ArrowLine(250,25)(300,25)\ArrowLine(300,25)(300,75)\ArrowLine(300,75)(250,75)
\DashArrowLine(300,25)(350,25){2}\DashArrowLine(350,75)(300,75){2}
\ArrowLine(350,25)(400,25)\ArrowLine(350,75)(350,25)\ArrowLine(400,75)(350,75)
\Text(310,50)[]{$j_L$}\Text(360,50)[]{$j_L$}
\Text(260,30)[]{$s_R$}
\Text(260,80)[]{$d_R$}\Text(390,80)[]{$s_R$}
\Text(390,30)[]{$d_R$} \Text(325,82)[]{$\chi^{-5/6}$}
\Text(325,32)[]{$\chi^{-5/6}$}

\end{picture}
\caption{\label{h1} Contributions to $K^0-\bar{K}^0$ mixing, where
the vertices are proportional to $h'^d$ and appropriate quark
mixing matrix elements. Similarly, we have the box diagrams with
the left-handed external quarks where $\phi$ gives also
contributions.} \ec
\end{figure}

The contributions to the $K^0-\bar{K}^0$ mixing come from the box
diagrams as shown in Fig. \ref{h1}.
After integrating out the heavy particles with a characteristic
mass scale $M$, the amplitude is proportional to $[(h')^4/(16\pi^2
M^2)](\bar{d}_{L,R}\ga_\mu s_{L,R})(\bar{d}_{L,R}\ga^\mu
s_{L,R})+h.c.,$ where $(h')^4$ is some product of $h'^d$, $h'^k$,
or $h'^j$ and appropriate quark mixing matrix elements. For $M$ of
order 100 GeV, it is negligible provided $h'< 10^{-2}$, which is
similar to \cite{pw}.

\begin{figure}[h] \bc
\begin{picture}(400,200)(0,0)
\ArrowLine(50,25)(90,25)\ArrowLine(90,25)(160,25)
\ArrowLine(160,25)(200,25) \DashArrowArcn(125,25)(35,180,0){2}
\Photon(149.75,49.75)(189.75,89.75){2}{6} \Text(60,30)[]{$b_R$}
\Text(190,30)[]{$s_R$} \Text(125,32)[]{$k_L$}
\Text(125,70)[]{$\chi^{+1/6}$} \Text(180,90)[]{$\gamma$}


\ArrowLine(250,25)(290,25)\ArrowLine(290,25)(360,25)
\ArrowLine(360,25)(400,25) \DashArrowArcn(325,25)(35,180,0){2}
\Photon(349.75,49.75)(389.75,89.75){2}{6} \Text(260,30)[]{$b_R$}
\Text(390,30)[]{$s_R$} \Text(325,32)[]{$j_L$}
\Text(325,70)[]{$\chi^{-5/6}$} \Text(380,90)[]{$\gamma$}


\ArrowLine(50,125)(90,125)\DashArrowLine(90,125)(160,125){2}
\ArrowLine(160,125)(200,125) \ArrowArcn(125,125)(35,180,0)
\Photon(149.75,149.75)(189.75,189.75){2}{6} \Text(60,130)[]{$b_R$}
\Text(190,130)[]{$s_R$} \Text(125,132)[]{$\chi^{+1/6}$}
\Text(125,170)[]{$k_L$} \Text(180,190)[]{$\gamma$}


\ArrowLine(250,125)(290,125)\DashArrowLine(290,125)(360,125){2}
\ArrowLine(360,125)(400,125) \ArrowArcn(325,125)(35,180,0)
\Photon(349.75,149.75)(389.75,189.75){2}{6}
\Text(260,130)[]{$b_R$} \Text(390,130)[]{$s_R$}
\Text(325,132)[]{$\chi^{-5/6}$} \Text(325,170)[]{$j_L$}
\Text(380,190)[]{$\gamma$}

\end{picture}
\caption{\label{h2} Contributions to $b\rightarrow s\gamma$, where
the vertices are proportional to $h'^d$ and appropriate quark
mixing matrix elements. Similarly, we have the diagrams with the
left-handed external quarks where $\phi$ gives also
contributions.} \ec
\end{figure}
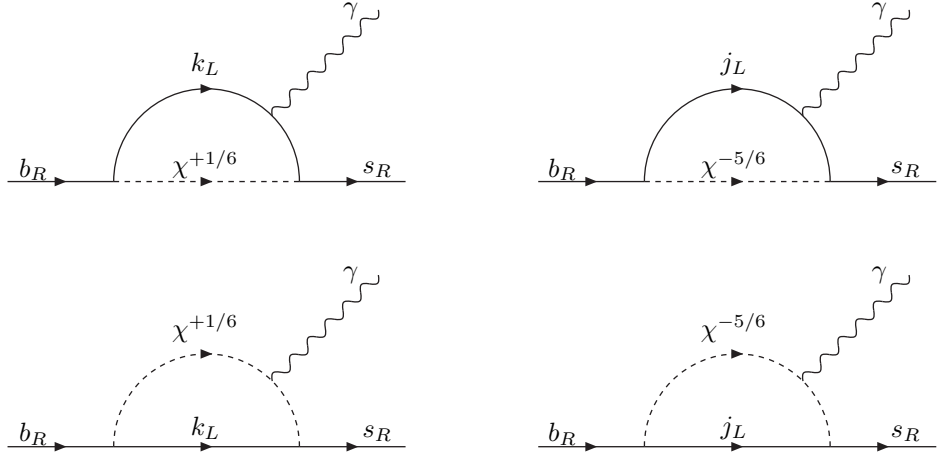

The contributions to $b\rightarrow s\gamma$ are given in Fig.
\ref{h2}.
In order to arrive at the branching ratio
$\mathrm{Br}(b\rightarrow s \gamma)$, we divide as usual the decay
width $\Ga(b\rightarrow s\gamma)$ by the theoretical expression
for the semileptonic decay width $\Gamma(b\rightarrow X e \nu)$
and multiply this ratio with the measured semileptonic branching
ratio $\mathrm{Br}(b\rightarrow X e \nu)\simeq 0.1$ \cite{pdg}:
\be \mathrm{Br}(b\rightarrow s \gamma)=\fr{\Ga(b\rightarrow
s\gamma)}{\Gamma(b\rightarrow X e \nu)}\mathrm{Br}(b\rightarrow X
e \nu).\label{br}\ee The semileptonic decay width (see, for
example, \cite{ghw}) is \be \Gamma(b\rightarrow X e \nu)\simeq
\fr{G^2_F
m^5_b|V_{cb}|^2}{192\pi^3}g(m_c/m_b)\left[1-(2/3\pi)\al_s(m_b)f(m_c/m_b)\right],\ee
where \bea g(x)&=&1-8x^2+8x^6-x^8-24 x^4 \ln x,\crn f(x)&=&
(\pi^2-31/4)(1-x)^2+3/2.\nn\eea The decay width in the model can
be evaluated as follows \be \Ga(b\rightarrow s\gamma) \simeq
\fr{\al m^5_b}{(12\pi)^4}\fr{|h'^{d*}_2 h'^d_3|^2}{M^4}.\ee Hence
eq. (\ref{br}) becomes: \be \mathrm{Br}(b\rightarrow s
\gamma)=\fr{\al G^{-2}_F|V_{cb}|^{-2}\mathrm{Br}(b\rightarrow X e
\nu)}{108\pi
g(m_c/m_b)\left[1-(2/3\pi)\al_s(m_b)f(m_c/m_b)\right]}\fr{|h'^{d*}_2
h'^d_3|^2}{M^4}.\ee Now taking $g(m_c/m_b)\simeq0.5$,
$f(m_c/m_b)\simeq2.5$, $\al_s(m_b)\simeq0.19$, $\al\simeq 1/135$,
$G_F\simeq10^{-5}\ \mathrm{GeV}^{-2}$, $V_{cb}\simeq0.04$, we have
\be \mathrm{Br}(b\rightarrow s \gamma)\simeq0.3 \left(\fr{100\
\mathrm{GeV}}{M}\right)^4|h'^{d*}_2 h'^d_3|^2.\ee The calculation
in the SM for this branching is in good agreement with the
experiments $\mathrm{Br}(b\rightarrow s \gamma)\simeq 3.55 \times
10^{-4}$ \cite{pdg}. In our case, the new physics does not give
contribution (i.e. does not contradict the SM result) if $M$ is of
order 100 GeV and $h'^d< 0.1$.

The flavor changing in the lepton sector is the same as in the
simplest seesaw extension of the SM. As such, it is to be noted
that the charged lepton flavor violations such as $\mu\rightarrow
e \gamma$ and $\mu \rightarrow 3e$ are highly suppressed
\cite{clh}.

\subsection{Stable anomalously-charged scalars}

As mentioned our model contains the scalars $\chi$ and $\phi$ with
the anomalous electric charges under which at least one of them
may be very long-lived. Indeed, due to the electric charge
conservation the lightest anomalously-charged scalar (assuming
$\chi^{\pm1/6}$) never decays, even if included quantum
corrections. This is similar to the supersymmetric models with
$R$-parity conservation there the lightest superparticle (LSP) can
be charged-scalars such as stau, stop, sbottom or the lightest
messenger in some scenario of the gauge-mediated supersymmetry
breaking \cite{digp} (see also \cite{cdm,cdm1}). In addition, our
proposal is not similar to the supersymmetric models there the
long-lived charged particles are the next-to-LSP (NLSP)
\cite{ditw}.

If our stable scalar $\chi$ is produced at a collider, it can
easily escape the detector. There is no simple way to measure its
lifetime (however, for searches of long-lived charged NLSPs, see
\cite{champ}). It is important to determine whether the lifetime
is indeed finite or if the particle is stable on cosmological
timescales. Its cosmological evolution can be obtained from the
Boltzmann equation: \be \fr{dn_\chi}{dt}=-3H n_\chi-\langle \sigma
v\rangle [n^2_\chi-(n^{eq}_\chi)^2],\ee where $n_\chi$ is the
number density, $n^{eq}_\chi$ is that of equilibrium, $H$ is the
Hubble parameter, and $\langle \sigma v\rangle$ is a
thermal-averaged annihilation cross section times their relative
velocity.

The dominant contributions to the annihilation cross section would
be model-dependent. In our case, it comes from the annihilation
processes into $\ga\ga$ and $Z\ga$. Moreover, due to an anomalous
electric charge $1/6$ for $\chi$ that its electromagnetic coupling
is weak, such annihilation processes are very rare. It follows
that the bounds of $\chi$ mass in our model may be much lower than
the other cases \cite{cdm1}. The typical lower bound on the mass
of such stable particles in various models is able to as low as 1
TeV \cite{cdm1}. Hence, the lower bound for the $\chi$ mass is
equal to $(1/6)^2\times 1$ TeV $\simeq$ 277 GeV (see, for example,
Chuzhoy and Kolb in \cite{cdm1}). This is in agreement with a
search by Byrne, Kolda and Regan in \cite{cdm1} for the lower mass
bound of the squarks and gluinos about 230 GeV. It is also in
agreement with searches by Smith {\it et al.} in \cite{cdm1} and
Yamagata {\it et al.} in \cite{cdm1}.

\subsection{Baryogenesis}

In this model the proton decay is discarded because the effective
operator $QQQL$ could not be generated with provided the model
particle content. Also, the neutrinoless double beta decay with
the effective operator $QQ\bar{Q}\bar{Q}LL$ is explicitly
suppressed. All these are the natural consequences due to the
local lepton-number and baryon-number conservations. In the
literature, the above processes are generally known to be
prevented up to the very high scale as of the grand unification
theories, where the lepton and baryon numbers may be violated. In
our model the status is different. Although these processes cannot
happen due to the gauge symmetries, there are still spontaneous
breaking of baryon-number and lepton-number due to the VEVs of
baryon-charged $S_B$ and lepton-charged $S_L$, respectively. And,
the scales for these breakings in principle may be arbitrary but
should be greater than the electroweak scale. These VEVs are just
the sources for associated phenomena to be happened, independent
of the explicit gauge symmetries.

As an example, the conservation of the lepton number and baryon
number is the reason why we cannot generate the baryon number
asymmetry as in the standard technics through explicit
lepton/baryon violation interactions as such baryogenesis via the
grand unification or baryogenesis via leptogenesis \cite{fuya}.
The baryogenesis in this model may appear either one of the
following cases: \ben
\item We may realize a baryon asymmetry via the
spontaneous symmetry breaking of the baryon number at the TeV
scale due to $S_B$ as well as the ordinary quark products
resulting from the decay of the unstable leptoquarks. The
procedure for achieving the excess of baryon number can closely
follow Harvey and Turner in Ref. \cite{baryogenesis}. The
calculations in Ref. \cite{pw1} in the models like ours have shown
that this is possible.
\item Baryogenesis via leptogenesis: Let us recall that the
neutrinos in our model gain the masses via the type I seesaw
mechanism. The difference here is that the Majorana masses for the
right-handed neutrinos are generated as a result from the
spontaneous breaking of the lepton number due to $S_L$. The
leptogenesis can be obtained via this source due to a nontrivial
vacuum of the lepton number. The VEV for $S_L$ must be very high
and the procedure for deriving an excess of lepton number, thus
the baryon number, is similar to \cite{fuya}. \een

A detailed calculation for these cases to be included in the
current work is out of the scope of this letter, and we will
devote it to a further publication.

\subsection{Other aspects}

Our model has two new gauge bosons as mentioned $Z'_L$ and $Z'_B$,
respectively gaining the masses via the lepton- and baryon-number
breaking VEVs. If they are much heavier than the electroweak scale
(particularly, this may exist in the second case of the
baryogenesis as mentioned) their contributions to the collider
phenomenology and affectations to the Higgs potential could be
negligible. However, as shown above, in the first case the
breaking of the baryon number at the TeV scale may be responsible
for the baryogenesis. Then the $Z'_B$ gets a mass in this scale.
We can also have a $Z'_L$ light as of $Z'_B$ (this may only happen
in the first case since the second case needs a very high scale of
lepton number breaking). The $Z'_L$ or $Z'_B$ can then contribute
to the known processes, e.g. $e^+ e^-\rightarrow Z'_L \rightarrow
\tau^+ \tau^-$ and $pp\rightarrow Z'_B \rightarrow t\bar{t}$, or
new processes to observe the lepton number violation
$pp\rightarrow \nu\nu$ provided that $Z'_L$ and $Z'_B$ mixing,
which can be used to search for (for a detailed evaluation, see
\cite{pw} and references therein).

At the LHC and ILC, we may have interesting processes due to the
decays of the leptoquarks. Here we focus on the case where the
leptoquarks decay into a stable scalar and a top quark. The
channels are (i) LHC: $pp\rightarrow \bar{j}j\rightarrow
\chi^{-1/6}\chi^{+1/6} \bar{t} t$, where the first process is
possibly mediated by $\gamma, Z,$ gluon, $Z'_B$ (if $Z'_B$ lies in
TeV scale with its large enough gauge coupling, this contribution
is also important) and (ii) ILC: $e^+e^-\rightarrow
\bar{j}j\rightarrow \chi^{-1/6}\chi^{+1/6} \bar{t} t$, where the
mediations of the first process may be $\gamma, Z$, $Z'_L$ (if
$Z'_L$ lies in TeV scale and its gauge coupling is large enough).
There is a possible portion of $\chi^{-1/6}\chi^{+1/6}$ fusion
into $\gamma\gamma$, $\gamma Z$, or a light fermion pair
$\bar{f}f$ via photon exchange but all these are very rare due to
a charge $1/6$ for $\chi$. The final state of those processes at
each stage may have a missing energy and a $\bar{t}t$ pair, which
is worth to looking for. Finally, we can have typical processes
with the stable scalar $\chi^{\pm 1/6}$ such as a direct channel
$e^+e^- \rightarrow \chi^{-1/6}\chi^{+1/6}$ at the ILC and
$\ga\ga\rightarrow \chi^{-1/6}\chi^{+1/6}$ at the photon-photon
collider. All the above processes are devoted to the forthcoming
experimental considerations.

Let us note that the similar processes can happen in Tevatron and
LEP, if the new particles are assumed to be light. The Tevatron
and LEP then provide constraints on the model. Otherwise, those
processes would been evaded if the new particles are much heavier
than the electroweak scale. In this case the model is explicitly
consistent with the effective theory, and we have a natural seesaw
mechanism for the neutrinos as well as the baryogenesis appearing
in the mentioned second case, that all are quite similar to the
standard context. Anyway, a constraint on the new physics at the
TeV scale using the existing colliders are worth, and a more
detailed analysis on the model's consequences as briefly mentioned
are needed. All these are large subjects out of scope of this
letter. We will study of these issues to be published elsewhere in
a near future.

In our model, with such a heavy leptoquark generation it is
well-established that the gluon gluon fusion cross section for the
SM Higgs is larger by a factor 9 \cite{ggf}. However, the new
results from CDF and D0 do not rule out our model when the Higgs
mass is $114\ \mathrm{GeV} < M_H <120\ \mathrm{GeV}$, or when
$M_H>200\ \mathrm{GeV}$ \cite{cdfdo}. Notice that for a large
mixing between $H$ and the singlets $S_{L}$ and $S_B$ (if these
scalars are assumed to be light enough) one can relax those
constraints. These are similarities to the former proposal
\cite{pw}.

\section{\label{conclus}Conclusions}

We have proposed a simple, predictive model of the gauged lepton
and baryon charges. All the anomalies were removed by the presence
of only the leptoquarks $j^{+1/2}$ and $k^{-1/2}$. The numbers of
the observed lepton and quark families have been proved to be
equal to the number of the fundamental colors, which is just
three. Let us note that in the standard model the number of
fermion families is left arbitrary, and thus fail to answer this
question. The right-handed neutrinos have been naturally existed
as a requirement of the gravity-anomaly cancelation. This is an
interesting feature in comparison with such particles as required
for the SO(10) grand unification. In contradiction to the previous
proposals \cite{pw}, the lepton sector in our model is minimal
since it only assumes the mentioned right-handed neutrinos.

The conservation of the lepton and baryon numbers will explicitly
prevent the proton decay and neutrinoless double beta decay, but
the spontaneous breaking of these charges will explain for
associated low-energy phenomena such as neutrino mass and
baryogenesis. Indeed, the small masses of the neutrinos have been
explained by the type I seesaw mechanism in similarity to the
simplest seesaw extension of the standard model. But, in our case
the Majorana masses for the right-handed neutrinos get naturally
generated as a result of spontaneous lepton-number breaking due to
the VEV of lepton-charged $S_L$ scalar, although this symmetry is
exact. A standard technic for the baryogenesis via leptogenesis is
therefore followed. In other case, the spontaneous breaking of
baryon number at TeV scale as well as the decay of the leptoquarks
may also generate the baryogenesis. We particularly stress that
the neutrino masses and leptogenesis as obtained in our model
provide an insight into the standard contexts.

The flavor changing neutral current processes such as
$K^0-\bar{K}^0$ mixing and $b\rightarrow s \gamma$ have been
considered and evaluated. Here the new physics constrained is in
natural consistency with the standard model. The model contains an
interesting stable scalar as associated with an anomalous electric
charge that can be very long-lived. This is similar to the
supersymmetric models where the LSP is electrically charged such
as stau, stop or sbottom. However, our scalar is different from
the ones mentioned because it has very weak electromagnetic
coupling. A naive evaluation has shown that the low bound on its
mass may be as low as 277 GeV.

If the scales of lepton- and baryon-number breaking are much
larger than the electroweak scale, the $S_L$ and $S_B$ would not
affect the Higgs potential,  and the gauge bosons associated with
these charges $Z'_{L}$ and $Z'_B$ do not contribute to the
collider phenomenology. However, if these scales are as low as
required in a case for the baryogenesis, they will take place.
Also, existing as an answer of consistency of the model the new
leptoquarks $j$, $k$ and the new scalars $\chi$ and $\phi$ with
the anomalous electric charges (where almost the new particles
fast decay, only one is the stable scalar as mentioned) have the
interesting phenomenologies in the colliders such as the LHC, ILC
and photon-photon collisions, which can be worth to search for. In
addition, such similar processes can also happen in the existing
accelerators such as the Tevatron and LEP, if the new particles
are light enough, which provide constraints on the model. A
detailed study on these issues are necessary to be published
elsewhere.

\section*{Acknowledgments}
This work is supported in part by the National Foundation for
Science and Technology Development of Vietnam (NAFOSTED).
\\[0.3cm]

\end{document}